\documentclass[twocolumn]{revtex4-1}
\usepackage{epsfig}
\usepackage{graphics}
\usepackage{amssymb}
\usepackage{textcomp}
\usepackage{amsmath}
\usepackage[dvipsnames]{xcolor}

\def\eq#1{(\ref{#1})}

\begin{document}

\title{Loop quantum cosmology on a brane via holography}

\author{C. A. S. Silva} 
\email{carlos.silva@ifpb.edu.br}
\affiliation{Instituto Federal de Educa\c{c}\~{a}o Ci\^{e}ncia e Tecnologia da Para\'{i}ba (IFPB),\\ Campus Campina Grande - Rua Tranquilino Coelho Lemos, 
671, Jardim Dinam\'{e}rica
I.}

\date{\today}

\begin{abstract}

In this letter, we demonstrate that Loop Quantum Cosmology equations, with a near vanishing late time cosmological constant, can be induced on a brane in a Randall-Sundrum II scenario, corresponding to the holographic dual of a strongly coupled string regime in the bulk. Such results can establish a possible
connection between Loop Quantum Gravity and string theory.

\end{abstract}


\maketitle

Loop Quantum Cosmology (LQC) \cite{Bojowald:2001xe}, 
which comes from Loop Quantum Gravity (LQG) , and braneworld cosmology \cite{Maartens:2003tw}, arising from string theory, 
suggest possible descriptions of the near Big-Bang physics,
mapping directions for several areas of cosmology and related research, and allowing make contact with observational activity \cite{Maartens:2003tw, Barrau:2013ula, Ashtekar:2015dja, Bojowald:2011hd}.

However, these two kinds of cosmology remain at variance until now, leading physics 
to sit on the fence about the beginning state of our universe.  
Even though, a possible duality between
them has been pointed \cite{Singh:2006sg}, they disagree about the role of 
the quantum corrections to Friedmann equations, which implies in a controversy
on the resolution of the Big Bang singularity.

On another standing point, the holographic principle has been established as guide principle to quantum gravity 
\cite{'tHooft:1993gx, Susskind:1994vu, Bousso:2002ju}.
In  particular, in braneworld context, such principle has been proposed in the lines of the AdS/CFT conjecture \cite{Maldacena:1997re}. The application of such conjecture to the Randall-Sundrum II braneworld scenario \cite{Randall:1999vf} has established that a gravitational theory in the AdS bulk is dual to a CFT with a UV cut-off, coupled to gravity on the brane \cite{Padilla:2002tg}.
Even though we do not know the precise nature of the gravitational theory on the brane, the intuition about it is that, as the bulk black hole emits Hawking radiation, it heats the brane to a finite temperature. Consequently, whatever the gravitational theory on the brane is, it should be hot, and have a non-zero energy density and pressure \cite{Padilla:2002tg}. In this case, one may wonder that the nature of the gravitational theory on the brane must be intrinsically related with the induced thermodynamics on it.

In this sense, an important result by Jacobson has been the derivation of
Einstein's field equations  from thermodynamics \cite{Jacobson:1995ab}.
The basic assumption behind this result is to consider that the Clausius relation, $dQ = TdS$, which connects entropy, temperature and heat, can be held through each spacetime point, for all the local Rindler causal horizon, by taking $T$ as the Unruh temperature and $dQ$ as the energy flux, as measured by an observer in an accelerated frame just inside the Rindler horizon.

Hitched in this discussion is the applicability of the holographic principle to cosmology.
Following the pioneer work by Fischller and Suskind \cite{Fischler:1998st}, a sort of scenarios has been proposed to establish the validity of the holographic principle in cosmological contexts, consisting in the so called Holographic Cosmology \cite{Fischler:1998st, Veneziano:1999ts, Bousso:1999xy, Bousso:1999cb, Easther:1999gk, Bak:1999hd, Kaloper:1999tt, Banks:2001px}. 
More recently, observational evidence has been pointed to the existence of a holographic phase in the earlier times of our universe \cite{Afshordi:2016dvb}. 
Among the versions of the holographic cosmology, Bak and Rey have argued that the holographic principle must be satisfied by the universe if one considers that its entropy must be associated with its apparent horizon \cite{Bak:1999hd}. In this context, the validity of the first law of thermodynamics can be proved, which has made possible to  derive the Friedmann equations of a FLRW universe \cite{Cai:2005ra, Gong:2007md} by the use of the Jacobson formalism \cite{Jacobson:1995ab}.

The applicability of such calculations, to obtain cosmological equations in four dimensions, has been proved also in the context of braneworld cosmology \cite{Ge:2007yu}, and recently in the context of LQC \cite{Silva:2015qna}. Consequently, one may wonder if such method could be one of the possible bridges between these
two kind of cosmologies.
To answer such question, we must note that a central point in both derivations consists in how to count the spacetime degrees of freedom in four dimensions. Particularly in the context of braneworld, such question is connected with what
$4D$ solution describing a black hole localized on the
brane is obtained. Such a solution would be given by a suitable slicing of a $5D$ accelerating
black hole metric, known as the C-metric in $4D$ \cite{Kinnersley:1970zw}.

Several proposals have been presented in the literature, where a induced $4D$ metric of the form

\begin{equation}
ds^{2} = - F(r)^{2}dt^{2} + G(r)^{2}dr^{2} + H(r)^{2}d\Omega^{2} \label{g-gorm} 
\end{equation}

\noindent has been used \cite{Chamblin:1999by, Dadhich:2000am, Casadio:2001jg, Visser:2002vg, Bronnikov:2003gx}. However some points remain to be clarified.

Among such points,  is that  related with the presence of a conformal ambiguity in the definition of the black hole metric on a brane, whose definition depends on  a conformal factor  \cite{Overduin:1998pn}. In this context, it arises the single answer question: “which choice of the conformal factor in the
four-dimensional metric reproduces Einstein’s gravity?"

The resolution of the conformal ambiguity problem is not the goal of the present article. However, we shall use it in order to discuss a further question, by  going beyond classical general relativity: is there a possible choice of the conformal factor in the metric induced on the brane that could generate the semiclassical LQC equations? In this way, the appearence of LQC equations on a brane could be used as a criteria in order to fix the conformal ambiguity problem.

In order to make the choice of the conformal factor more natural in a braneworld scenario, we shall observe some AdS/CFT  lessons.
At first, we note that all the approaches to find out a black hole solution on a brane  
have the common assumption that the radial coordinate must be fixed by the "`area gauge"' $H(r) = r$, in
the metric. Consequently, the area of the $2$-spheres surrounding the black hole behaves as
$A(r) = 4\pi r^{2}$, increasing monotonically
between the horizon and the spacelike infinity.
However, a different perspective has been driven by the AdS/CFT conjecture \cite{Maldacena:1997re}, where it has been proposed that black hole solutions localized on the brane 
given as solutions of the classical bulk equations in $AdS_{D+1}$ with the
brane boundary conditions, correspond to quantum-corrected black holes in
$D$ dimensions, rather than classical ones \cite{Emparan:2002px}.
In this context, it has been demonstrated that a static
quantum black hole solution must have a singular horizon when one takes into account
the dual description of a CFT with a cutoff living on the brane.

Based on such results, Gregory et al \cite{Gregory:2004vt} have argued that 
the area function $A(r)$ of the $2$-
spheres, in the geometry induced on the brane, must be considered to be not monotonic. 
The main reason to such assumption comes from
the presumed higher dimensional $C$-metric. Such a metric would consist of an accelerating
black hole being "`pulled" by a string. Since the appropriate higher dimensional metric
for a Poincar\'{e} invariant string has a turning point in the area function, and the
"horizon" is singular, the braneworld black hole horizon must also becomes
singular, and a turning point must occur in the area function of the $4D$ braneworld black hole metric.
In front of such arguments, it is interesting to investigate more general 
ansatze than $H = r$.


We can utilize such discussions in order to address the problem of conformal ambiguity by noting  that a metric of the form \eq{g-gorm}, for the case where  $H(r) \neq r$, can be find out by taking a conformal transformation of a metric written in the area
gauge $d{s}_{ag}^{2}$,

\begin{equation}
ds^{2} = \frac{H^2}{r^2} d{s}_{ag}^{2} \; ,  \label{resc-eq}
\end{equation}

\noindent by requiring that $H/r$ is a
twice differentiable function of the spacetime coordinates, and  $0 < H/r <  \infty$.

The expression \eq{resc-eq} makes possible to attend the Gregory requirement by a suitable choice of the conformal factor. We can  choose, in particular:

\begin{equation}
H^2(r) = \Big(r^{2}+
\frac{L^{4}}{r^{2}}\Big) \label{conformal-factor} \; , 
\end{equation}

\noindent where $L$ is an arbitrary parameter, which we shall consider to be of the order of the Planck length. For $L \rightarrow 0$, we recover the area gauge. Such conformal factor has been used in \cite{Modesto:2016max, Bambi:2016wdn, Bambi:2016yne}  to solve the  Schwarschild black hole singularity. 

As we shall see, the conformal factor \eq{conformal-factor} will correspond to a suitable choice in order to obtain the LQC equations on the brane. 
However, one could test other choices. If it is possible to find out another, the problem of conformal ambiguity is not completely solved in the present scenario. If is not possible to find out it,  the conformal ambiguity problem could be solved, if one consider that the appearence of LQC equations can be used as a sufficient criteria in order to choose the correct form of the black hole metric induced on the brane.

An important consequence of the conformal relationship between the metrics $ds^{2}$ and $d{s}_{ag}^{2}$ is that the way how the black hole temperature and entropy are related with the black hole mass is the same
in both spacetimes \cite{Bambi:2016yne}. Particularly, in concern to the $d{s}_{ag}^{2}$ spacetime, we shall not be interested in its specific form. However, we shall done the minimal assumption that such spacetime obeys the usual Bekenstein-Hawking entropy-area relation $S = \frac{A}{4}$, in four dimensions \cite{Bekenstein:1973ur}. (Such assumption can be satisfied for general class of braneworld black holes 
\cite{Abdalla:2006qj}.)

By considering the factor \eq{conformal-factor}, a new radial coordinate in the $ds^{2}$ spacetime will be defined as

\begin{equation}
R = \Big(r^{2}+
\frac{L^{4}}{r^{2}}\Big)^{1/2} \label{phys-rad}\; ,
\end{equation}

\noindent where $r$ is the radial coordinate in the spacetime described by $d{s}_{ag}^{2}$. 
Particularly, the value of $R$ associated with the black hole
event horizon will be

\begin{equation}
R_{EH} = \Big(r_{+}^{2}+
\frac{L^{4}}{r_{+}^{2}}\Big)^{1/2} \; ,
\end{equation}

\noindent where $r_{+}$ is the black hole event horizon radius in the $d{s}_{ag}^{2}$ spacetime.

Since the event horizon surface area $A$ changes by the conformal transformation, the black hole entropy-area relation
must be modified. In fact, the rescaled event horizon area will be given by

\begin{equation}
A = 4\pi\Big(r_{+}^{2}+
\frac{L^{4}}{r_{+}^{2}}\Big)\;\; , \label{r-area}
\end{equation}

\noindent and the black hole entropy will be expressed in terms of the area \eq{r-area} as

\begin{equation}
S  = \frac{\sqrt{A^{2} - A_{L}^{2}}}{4} + \ensuremath{\mathcal{O}(\geq A_{L}^{6}})\;\;\; , \label{entropy-area}
\end{equation}

\noindent where $A_{L} = 4\sqrt{2\pi} L^{2}$.
Such expression also appears in the context of loop quantum black holes \cite{Modesto:2009ve}, and have been used to derive the LQC dynamical equations from the holographic principle \cite{Silva:2015qna}.

Anchored in such facts,  in order to find out the equations for the cosmological evolution of the universe,  we shall apply the point of view by Bak and Rey \cite{Bak:1999hd} for holographic cosmology,  further developed by Cai and Kim by the use of the Jacobson's formalism \cite{Cai:2005ra}, and adapted by Ge to the braneworld scenario \cite{Ge:2007yu}.

In this context,  the energy flux through the apparent horizon is measured by a Kodama observer inside it \cite{Hayward:1997jp, Hayward:1994bu, Racz:2005pm, Ge:2007yu, Cai:2008gw}. The temperature measured by such observer is given by \cite{Cai:2008gw},

\begin{equation}
T = (2k_{B}\pi\tilde{r}_{A})^{-1}\;\;, \label{kodama-temperature}
\end{equation} 

\noindent where $\tilde{r}_{A}$ is the physical radius associated with the apparent horizon.

We can locally introduce a FLRW metric induced on the brane

\begin{eqnarray}
ds^{2} &=& h_{ab}dx^{a}dx^{b} + \tilde{r}^{2}d\Omega^{2}_{2}\label{frw-metric} \; ,
\end{eqnarray}

\noindent where $h_{ab} = diag (-1, a^{2}/(1- kr^{2}))$, $\tilde{r} = a(t)r$, and

\begin{equation}
\tilde{r}_{A} = \frac{1}{\sqrt{H^{2} + k/a^{2}}} \label{ap.hor-rad}\;. 
\end{equation}

Taking the path by Ge \cite{Ge:2007yu}, let us consider that the energy-momentum tensor $T_{\mu\nu}$ of the matter-energy contend 
in the universe is written as the ones for a perfect
fluid:

\begin{equation}
T_{\mu\nu} = (\rho + p)U_{\mu}U_{\nu} + p g_{\mu\nu}\;.
\end{equation}

\noindent From the energy conservation law, we obtain the continuity equation

\begin{equation}
\dot{\rho} + 3H(\rho + p) = 0\;. \label{continuity-eq}
\end{equation}

Now, let us write the expressions for the work density $W$ and the energy-supply vector $\psi$:

\begin{equation}
W = -\frac{1}{2}T^{ab}h_{ab}\; \hspace{3mm} \textrm{and} \hspace{3mm}
\psi_{a} = T^{b}_{a}\partial_{b}\tilde{r} + W\partial_{a}\tilde{r} \; .
\end{equation}

\noindent In a  Randall-Sundrum II braneworld scenario, we have

\begin{equation}
W = \frac{1}{2}(\rho - p - \frac{1}{6}\frac{k_{5}^{4}}{k_{4}^{2}}\rho p)\;;
\end{equation}

\noindent and

\begin{equation}
\psi_{t} =  - \frac{1}{2}(\rho + p)H\tilde{r} - \frac{1}{12}\frac{k_{5}^{4}}{k_{4}^{2}}\rho(\rho+ p)H\tilde{r}\; ,
\end{equation}

\begin{equation}
\psi_{r} =  \frac{1}{2}(\rho + p)a + \frac{1}{12}\frac{k_{5}^{4}}{k_{4}^{2}}\rho(\rho+ p)a \; ,
\end{equation}

\noindent where $k_{n}^2 = 8\pi G_{n}$, with $G_{n}$ consisting in the Newton's gravitational constant in n-dimensions.

By the use of the relations above, we can find out the expression for the amount of energy that goes through 
each apparent horizon during a time $dt$ as \cite{Cai:2005ra}

\begin{eqnarray}
\delta Q = A(\rho+p)\tilde{r}H\Big(1+ \frac{\rho}{\sigma}\Big)dt \; ,
\end{eqnarray}

\noindent where $A = 4\pi\tilde{r}^{2}_{A}$, and $\sigma = 6k_{4}^{2}/k^4_{5}$ is the brane tension.

In concern to the entropy of the apparent horizon on the brane, considering that the universe obeys the holographic principle,  it must be given by the entropy associated with the induced black hole solution on the brane. In this way, following the above discussions based on the AdS/CFT conjecture,  the apparent horizon entropy will not obey the area formula, in its classical form,  but it must be a quantum corrected one. Another important detail is that, in a general  braneworld scenario, it is expected that  the apparent horizon entropy is given by a $5$-dimensional relation, which contains contributions from the bulk \cite{Cai:2006pa}. However, as has been argued by Ge \cite{Ge:2007yu}, in the application of the Jacobson formalism, where  heat exchanges can be considered as approximately isometric,  the five-dimensional contributions to the apparent horizon entropy are dropped, and it  becomes to be given by a $4$-dimensional  formula. 
Following such discussions, we shall consider that the entropy of the apparent horizon on the brane is given by the $4$-dimensional quantum corrected 
entropy-area relation 
\eq{entropy-area}. 

In this case, by the use of the Clausius relation
$dQ = TdS$, and the temperature \eq{kodama-temperature}, 
we obtain

\begin{equation}
\dot{H} - \frac{k}{a^{2}} =   4\pi \frac{\sqrt{A^{2}-A_{L}^{2}}}{A}(\rho + p)\Big(1+ \frac{\rho}{\sigma}\Big) 
\;.
\label{friedmann1}
\end{equation}

By the use of the continuity Eq. \eq{continuity-eq}, we get

\begin{equation}
\frac{8\pi}{3}\frac{d\rho}{dt}\Big(1+ \frac{\rho}{\sigma}\Big)  = 
\frac{A}{\sqrt{A^{2} - A_{L}^{2}}} \frac{d(H^{2}+k/a^{2})}{dt} \; ,
\end{equation}

\noindent which by integration give us

\begin{equation}
H^2 + \frac{k}{a^2} = \frac{4\pi}{A_{L}}\cos(\Theta) \;, \label{friedmann2}
\end{equation}

\noindent where $\Theta = \pm\Big[\frac{2A_{L}}{3}\rho\Big(1+ \frac{1}{2}\frac{\rho}{\sigma}\Big) - \alpha\Big]$, and 
$\alpha$ is a phase 
constant.

In the Eq. \eq{friedmann2}, the effective density term 
in the form of a harmonic function of the classical density bring us a scenario where a quantum bounce takes the place of the Big Bang initial 
singularity when the universe density assumes a critical value, in a different way from usual braneworld cosmology. 

The phase constant $\alpha$ in the equation \eq{friedmann2} will depend on the initial conditions of the universe. Among such initial conditions, is the universe size at its beginning which corresponds to the minimal value of the universe apparent horizon radius, given by the equations \eq{ap.hor-rad} and \eq{friedmann2} as $\sqrt{A_{L}/4\pi}$. In the limit where $A_{L} \rightarrow 0$, which will correspond to the classical Big Bang singularity at the beginning of the universe, we obtain from the equation \eq{friedmann2}, $\cos(\pm\alpha) \rightarrow 0$, i.e., $\alpha \rightarrow \pi/2$. In this limit, we recover the usual braneworld cosmology.

By expanding the right-hand side of the equation \eq{friedmann2}, we get

\begin{equation}
H^{2} + \frac{k}{a^{2}} = A(\alpha)\rho^{2} + B(\alpha)\rho + C(\alpha)  \; , \label{friedmann2.2}
\end{equation}

\noindent where

\begin{eqnarray}
A(\alpha) = &&\frac{4\pi}{9}\Big(\frac{3\sin{(\alpha)}}{\sigma} - 2A_{L}\cos{(\alpha)}\Big) , \nonumber \\
B(\alpha) = &&\frac{8\pi}{3}\sin{(\alpha)}, \nonumber \\
C(\alpha) = &&\frac{4\pi}{A_{L}}\cos(\alpha)\; . \label{coefficients}
\end{eqnarray}

\noindent Here, we have  discarded the terms of higher orders in  $A_{L}$, as done in usual LQC \cite{Taveras:2008ke}.

The function $C(\alpha)$ is related with the late time cosmological constant, i.e, the value
of the cosmological constant when $\rho \sim 0$, given by

\begin{equation}
\Lambda_{LT} \sim \frac{3}{2A_{L}}\cos{(\alpha)}. \label{late-time}
\end{equation}

The Eq. \eq{friedmann2.2} can be written in the form

\begin{eqnarray}
H^{2} + \frac{k}{a^{2}} = \frac{8\pi}{3}\rho_{tot}\Big(1 - \frac{\rho_{tot}}{\rho_{c}}\Big) \label{lqc-eq}\;\;,
\end{eqnarray}

\noindent where $\rho_{tot} = \rho + \frac{\Lambda}{8\pi}$, with $\Lambda$ as a cosmological constant.

The Raychaudhuri equation can also be obtained 

\begin{equation}
\dot{H} - \frac{k}{a^{2}}  = -4\pi(\rho_{tot}+p_{tot})\Big(1 - \frac{2\rho_{tot}}{\rho_{c}}\Big) \;\; , \label{raychaudhuri-eq}
\end{equation}

\noindent where $p_{tot} = p - \frac{\Lambda}{8\pi}$.

Note that above the sign of $\rho_{c}$ has been included in its definition in a way that, even though the universe critical density  is a positive definite quantity,  $\rho_{c}$ can appears in our results as positive or negative. (In according with the equations \eq{lqc-eq} and \eq{raychaudhuri-eq}, positive for LQC, and negative for braneworld).

From \eq{friedmann2.2}, \eq{coefficients} and \eq{lqc-eq}, we must have 

\begin{equation}
\rho_{c}^{-1} = \frac{1}{6}\Big(2A_{L}\cos{(\alpha)} - \frac{3\sin{(\alpha)}}{\sigma}\Big)\;\;, \label{eq-i}
\end{equation}
\begin{equation}
1 - \frac{2\tilde{\Lambda}}{\rho_{c}} = \sin{(\alpha)}\;\;, \label{eq-ii}
\end{equation}
\begin{equation}
\Lambda_{LT} = \tilde{\Lambda}\Big(1-\frac{\tilde{\Lambda}}{\rho_{c}}\Big) \;\;. \label{eq-iii}
\end{equation}

\noindent where $\tilde{\Lambda} = \frac{\Lambda}{8\pi}$.

We shall fix the cosmological constant to be Planckian, as foreseen by the Standard Model \cite{Padilla:2015aaa}. However,  in order adjust the late time cosmological constant to observations ($\tilde{\Lambda}_{LT} \sim 10^{-122}$, in Planck units \cite{Riess:1998cb, Perlmutter:1998np}), we must have, from \eq{eq-ii} and \eq{eq-iii},  $\sin{(\alpha)} \sim -1$, and $\cos{(\alpha)} \sim 0$.
In this case, from \eq{eq-i},

\begin{equation}
\frac{1}{\rho_{c}} \sim \frac{1}{2\sigma} > 0 \;\;. \label{r-sigma}
\end{equation}

\noindent Note that $\sigma$ is a positive quantity in a Randall-Sundrum II scenario \cite{Randall:1999vf}, as well as $\rho_{c}$ in LQC .

The equations \eq{lqc-eq}, and \eq{raychaudhuri-eq} will correspond to the semiclassical LQC equations, where the universe critical density is related with the brane tension.
 Note that the appearance of LQC equations turned to be a necessary condition  in order to have a near vanishing late-time cosmological constant.
 
The result in the equation \eq{r-sigma}  is not completely strange in the braneworld context. LQC-like equations in the form of the equations
\eq{lqc-eq} and \eq{raychaudhuri-eq}, with $\rho_{c} \sim 2\sigma$, has been already obtained in the literature, under the assumption of the existence of an extra temporal dimension \cite{Shtanov:2002mb}. However, the assumption of an extra temporal dimension brings up some problems, as
the presence of tachyonic modes, and the violation of causality and unitarity  \cite{Dvali1999}.

A crucial consequence from the equations \eq{lqc-eq} and \eq{raychaudhuri-eq} will be a difference imposed to the Hamiltonian structure of the gravitational theory induced on the brane. In fact, we have a result by Singh and Soni \cite{Singh:2015jus} that, from a Raychaudhuri equation in the form of the equation \eq{raychaudhuri-eq}, the following Hamiltonian can be obtained, for the case of a plane universe:

\begin{equation}
\mathcal{H} = \frac{-3V}{16\pi  \lambda^2}(1-\cos{(p\sqrt{\Delta})})\;, \label{lqc-hamiltonian}
\end{equation}
\noindent In the equation above, $p$ is the conjugate momentum  to the volume $V$. 
Moreover, $\lambda = (3/(32\pi G\rho_{c}))^{1/2}$ and $\sqrt{\Delta} = 8 \pi \lambda$, where both possess dimensions of length. In this scenario, we have

\begin{equation}
\rho_{c} = 6/\Delta \label{r-delta} \;.
\end{equation}

Singh and Soni has argued that the canonical
structure of a theory described by the Hamiltonian above is that related with a modified phase space for gravity which consists in
a polymerized phase space. Consequently, the underlying theory can be quantized in a background independent way, i.e., by the use of the polymer quantization.

In fact, if we promote the Hamiltonian \eq{lqc-hamiltonian} to a quantum operator, we obtain that it will inevitably  be written in terms of a shift operator $\widehat{W}(\sqrt{\Delta})$, in the place of the momentum operator, as occurs in LQC:

\begin{eqnarray}
\widehat{\mathcal{H}} &=& \frac{-3V}{16\pi G \lambda^2}(1-\widehat{\cos{(\sqrt{p\Delta})})} \nonumber \\
                      &=& \frac{-3V}{32\pi G \lambda^2}(2 - \widehat{W}(\sqrt{\Delta}) - \widehat{W}(-\sqrt{\Delta})) \;\; , 
\end{eqnarray}

\noindent where $\widehat{W}(\sqrt{\Delta}) = \widehat{e^{i\sqrt{\Delta} p}}$. Such shift operators are equivalent to holonomies in the full LQG \cite{Bojowald:2015iga}.
 
Consequently, a new phase space structure is introduced, where for a state $\psi_{x} = <p\!\!\mid\!\! x> = e^{ipx}$ in the momentum representation, we have

\begin{equation}
\widehat{W}(\sqrt{\Delta})\psi_{x} = e^{i\sqrt{\Delta} p}e^{i px} = e^{i(x + \sqrt{\Delta}) p} = \psi_{x+\sqrt{\Delta}}\;\; , \label{v-equation} 
\end{equation}

\noindent {in a way that the application of the operator $\widehat{W}(\beta)$ will correspond to a finite displacement equals 
to $\sqrt{\Delta}$.  
In this way, the presence of a shift operator in the Hamiltonian introduces the definition of a lattice $\gamma_{\sqrt{\Delta}}$ on the configuration space as $\gamma_{\sqrt{\Delta}} = \{x \in \mathbb{R}  \mid x = n\sqrt{\Delta}, \forall n \in \mathbb{Z} \}$, which give us a discreteness of the position $x$, with discreteness parameter $\sqrt{\Delta}$. It corresponds to the polymerized phase space used in LQG quantization techniques \cite{Bojowald:2015iga}. 
In the present context, the shift operator will change the scale factor $a$ (or some
function of it, such as an area or volume) by a smallest discrete increment, and the differential
equation for the evolution of the universe, will turn into a difference equation.

Interestingly, from the equations \eq{r-sigma} and \eq{r-delta}, we have that $\Delta$, which will correspond to the area gap in the discrete geometry defined on the brane, will be such that:

\begin{equation}
\Delta \sim \frac{1}{\sigma}.
\end{equation}

The result above allows us to relate the discrete geometry, modeled by LQG on the brane, with a string theory in the bulk. It is because the brane tension is connected with the string couplings \cite{Becker:2007zj}. In particular, we have $g_{s} \sim 1/\sigma$, where $g_{s}$ corresponds to the closed string coupling.
In this way,  we find

\begin{equation}
\Delta  \sim   g_{s} .
\end{equation}

\noindent Such result points to a possible intriguing link between LQG and string theory, where a strongly coupled string regime in the bulk will correspond to a discrete area spectrum on the brane described by LQG. For a weakly coupled string regime, the area spectrum goes to be continuous. Moreover, the appearance of LQC equations in a braneworld scenario as a necessary condition  to have a near vanishing late-time cosmological constant seems to be a light, turned on along the bridge established here between string theory and LQG, on the cosmological constant problem \cite{Padilla:2015aaa}. Further investigations on this issue must be done. At last, in the present context, both the universe critical density and the area gap are defined in terms of the string coupling, that is determined dynamically \cite{Zwiebach:2004tj}. It avoids the problem of the Immirzi ambiguity \cite{Rovelli:1997na}, and  matches the concern by several authors that  the Barbero-Immirzi parameter must be determined dynamically  \cite{Jacobson:2007uj, Taveras:2008yf, Mercuri:2009zi}.







\begin{thebibliography}{30}



  
\bibitem{Bojowald:2001xe}
  M.~Bojowald,
  Phys.\ Rev.\ Lett.\  {\bf 86}  5227 (2001).
   [gr-qc/0102069].

\bibitem{Maartens:2003tw}
  R.~Maartens,
  Living Rev.\ Rel.\  {\bf 7}  7 (2004).
  [gr-qc/0312059].

\bibitem{Barrau:2013ula}
 A.~Barrau, T.~Cailleteau, J.~Grain and J.~Mielczarek,
  Class.\ Quant.\ Grav.\  {\bf 31}  053001 (2014).
  
 

\bibitem{Ashtekar:2015dja}
 A.~Ashtekar and A.~Barrau,
  Class.\ Quant.\ Grav.\  {\bf 32}  no.23,  234001 (2015).
  [gr-qc/1504.07559].

\bibitem{Bojowald:2011hd}
  M.~Bojowald, G.~Calcagni and S.~Tsujikawa,
  Phys.\ Rev.\ Lett.\  {\bf 107}  211302 (2011).
  [astro-ph.CO/1101.5391].


\bibitem{Singh:2006sg}
  P.~Singh,
  Phys.\ Rev.\ D {\bf 73}  063508 (2006).
 [gr-qc/0603043].




	


	
	
	
	

	
	
	

	
	
  \bibitem{'tHooft:1993gx}
  G.~'t Hooft,
  Salamfest 1993:0284-296
  [gr-qc/9310026].
  
  \bibitem{Susskind:1994vu} 
  L.~Susskind,
  J.\ Math.\ Phys.\  {\bf 36}  6377 (1995).
  
   \bibitem{Bousso:2002ju}
  R.~Bousso,
  Rev.\ Mod.\ Phys.\  {\bf 74}  825 (2002).
	
	\bibitem{Maldacena:1997re}
  J.~M.~Maldacena,
  Int.\ J.\ Theor.\ Phys.\  {\bf 38} (1999) 1113
   [Adv.\ Theor.\ Math.\ Phys.\  {\bf 2}  231 (1998)].

\bibitem{Randall:1999vf} 
  L.~Randall and R.~Sundrum,
  Phys.\ Rev.\ Lett.\  {\bf 83}, 4690 (1999)
  [hep-th/9906064].
	
.
	
	\bibitem{Padilla:2002tg}
  A.~Padilla,
 [ hep-th/0210217].
	
	

 \bibitem{Jacobson:1995ab}
  T.~Jacobson,
  Phys.\ Rev.\ Lett.\  75  1260 (1995).
	
	
	

	\bibitem{Fischler:1998st}
  W.~Fischler and L.~Susskind,
  [hep-th/9806039].
	

	
	\bibitem{Veneziano:1999ts}
  G.~Veneziano,
  Phys.\ Lett.\ B {\bf 454}  22 (1999).

\bibitem{Easther:1999gk}
R.~Easther and D.~A.~Lowe,
Phys.\ Rev.\ Lett.\  {\bf 82}  4967 (1999).

\bibitem{Bak:1999hd}
  D.~Bak and S.~J.~Rey,
  Class.\ Quant.\ Grav.\  {\bf 17}  L83 (2000).
	
	

	
	\bibitem{Bousso:1999xy}
  R.~Bousso,
  JHEP {\bf 9907}  004 (1999).
	
	\bibitem{Bousso:1999cb}
  R.~Bousso,
  JHEP {\bf 9906}  028 (1999).
	
	

\bibitem{Banks:2001px}
  T.~Banks and W.~Fischler,
  hep-th/0111142.
	
	\bibitem{Kaloper:1999tt}
N.~Kaloper and A.~D.~Linde,
Phys.\ Rev.\ D {\bf 60}  103509 (1999).

  

\bibitem{Afshordi:2016dvb}
  N.~Afshordi, C.~Coriano, L.~Delle Rose, E.~Gould and K.~Skenderis,
  Phys.\ Rev.\ Lett.\  {\bf 118}  no.4,  041301 (2017).
	
	
	
	\bibitem{Cai:2005ra}
  R.~G.~Cai and S.~P.~Kim,
  JHEP  0502  050 (2005).

  \bibitem{Gong:2007md}
  Y.~Gong and A.~Wang,
  Phys.\ Rev.\ Lett.\  {\bf 99}  211301 (2007).
	
\bibitem{Ge:2007yu}
  X.~H.~Ge,
  Phys.\ Lett.\ B {\bf 651}  49 (2007).

\bibitem{Silva:2015qna}
  C.~A.~S.~Silva,
  Eur.\ Phys.\ J.\ C {\bf 78}  no.5,  409 (2018).
	

	

	\bibitem{Kinnersley:1970zw}
  W.~Kinnersley and M.~Walker,
  Phys.\ Rev.\ D {\bf 2}  1359 (1970).
	

	
	
	
	
	\bibitem{Chamblin:1999by}
  A.~Chamblin, S.~W.~Hawking and H.~S.~Reall,
  Phys.\ Rev.\ D {\bf 61}  065007 (2000).
	
	\bibitem{Dadhich:2000am}
  N.~Dadhich, R.~Maartens, P.~Papadopoulos and V.~Rezania,
  Phys.\ Lett.\ B {\bf 487}  1 (2000).
	
	\bibitem{Casadio:2001jg}
  R.~Casadio, A.~Fabbri and L.~Mazzacurati,
  Phys.\ Rev.\ D {\bf 65}  084040 (2002).
	

	
	\bibitem{Visser:2002vg}
  M.~Visser and D.~L.~Wiltshire,
  Phys.\ Rev.\ D {\bf 67}  104004 (2003).
	
	\bibitem{Bronnikov:2003gx} 
  K.~A.~Bronnikov, V.~N.~Melnikov and H.~Dehnen,
  Phys.\ Rev.\ D {\bf 68}, 024025 (2003)
	
	
\bibitem{Overduin:1998pn} 
  J.~M.~Overduin and P.~S.~Wesson,
  Phys.\ Rept.\  {\bf 283}, 303 (1997)
	

\bibitem{Emparan:2002px}
  R.~Emparan, A.~Fabbri and N.~Kaloper,
  JHEP {\bf 0208}  043 (2002).

\bibitem{Gregory:2004vt}
  R.~Gregory, R.~Whisker, K.~Beckwith and C.~Done,
  JCAP {\bf 0410}  013 (2004).
  [hep-th/0406252].
	
	
	
	

	
\bibitem{Bambi:2016yne}
  C.~Bambi, L.~Modesto, S.~Porey and L.~Rachwal,
  JCAP {\bf 1709}  no.09,  033 (2017).

	\bibitem{Modesto:2016max}
  L.~Modesto and L.~Rachwal,
  arXiv:1605.04173 [hep-th].
	
	\bibitem{Bambi:2016wdn}
  C.~Bambi, L.~Modesto and L.~Rachwal,
  JCAP {\bf 1705}  no.05,  003 (2017).
	
	
	
\bibitem{Bekenstein:1973ur}
  J.~D.~Bekenstein,
  Phys.\ Rev.\ D {\bf 7}  2333 (1973)
	
	
	\bibitem{Abdalla:2006qj} 
  E.~Abdalla, B.~Cuadros-Melgar, A.~B.~Pavan and C.~Molina,
  Nucl.\ Phys.\ B {\bf 752}, 40 (2006)

\bibitem{Modesto:2009ve}
  L.~Modesto and I.~Premont-Schwarz,
  Phys.\ Rev.\ D {\bf 80}  064041 (2009).

\bibitem{Cai:2008gw}
  R.~G.~Cai, L.~M.~Cao and Y.~P.~Hu,
  Class.\ Quant.\ Grav.\  26  155018 (2009).



	
	
		
	
	

	\bibitem{Hayward:1997jp}
  S.~A.~Hayward,
  Class.\ Quant.\ Grav.\  {\bf 15}  3147 (1998).
	
	
	
	\bibitem{Hayward:1994bu}
  S.~A.~Hayward,
  Phys.\ Rev.\ D {\bf 53}  1938 (1996).
	
	\bibitem{Racz:2005pm}
  I.~Racz,
  Class.\ Quant.\ Grav.\  {\bf 23}  115 (2006).
	
		\bibitem{Cai:2006pa}
  R.~G.~Cai and L.~M.~Cao,
  Nucl.\ Phys.\ B {\bf 785}  135 (2007).
	
	
	
	
	


	
	
	
	\bibitem{Taveras:2008ke}
  V.~Taveras,
  Phys.\ Rev.\ D {\bf 78}  064072 (2008).
  [gr-qc/0807.3325].


	
\bibitem{Padilla:2015aaa}
  A.~Padilla,
  [hep-th/1502.05296].
	
	\bibitem{Riess:1998cb}
  A.~G.~Riess {\it et al.} [Supernova Search Team],
  Astron.\ J.\  {\bf 116}  1009 (1998).
	
	\bibitem{Perlmutter:1998np} 
  S.~Perlmutter {\it et al.} [Supernova Cosmology Project Collaboration],
  Astrophys.\ J.\  {\bf 517}  565 (1999).

\bibitem{Shtanov:2002mb}
 Y.~Shtanov and V.~Sahni,
Phys.\ Lett.\ B {\bf 557} (2003) 1
 [gr-qc/0208047].

\bibitem{Dvali1999}
 G.R. Dvali, G. Gabadadze, G. Senjanovic, in " Many Faces of the Superworld:Yuri Golfand
Memorial Volume" pp 525-532, Eds. Y. Golfand, M. Shifman, M.A. Shifman (World Scientific,
1999).
	
\bibitem{Singh:2015jus}
  P.~Singh and S.~K.~Soni,
  Class.\ Quant.\ Grav.\  {\bf 33}  no.12,  125001 (2016).
	
	


\bibitem{Bojowald:2015iga} 
  M.~Bojowald,
  Rept.\ Prog.\ Phys.\  {\bf 78}, 023901 (2015)
	
\bibitem{Becker:2007zj}
  K.~Becker, M.~Becker and J.~H.~Schwarz, New York, Cambridge University Press (2007).

\bibitem{Zwiebach:2004tj} 
  B.~Zwiebach,
  Cambridge, UK: Univ. Pr. (2009) 673 p

\bibitem{Rovelli:1997na} 
  C.~Rovelli and T.~Thiemann,
  Phys.\ Rev.\ D {\bf 57}, 1009 (1998)

\bibitem{Jacobson:2007uj}
 T.~Jacobson,
  Class.\ Quant.\ Grav.\  {\bf 24}  4875 (2007).
  [gr-qc/0707.4026].


\bibitem{Taveras:2008yf}
  V.~Taveras and N.~Yunes,
  Phys.\ Rev.\ D {\bf 78}  064070 (2008).
  [gr-qc/:0807.2652].
	
	\bibitem{Mercuri:2009zi}
  S.~Mercuri,
  Phys.\ Rev.\ Lett.\  {\bf 103}  081302 (2009).
 [gr-qc/0902.2764].
	

	
	
	
	








\end{thebibliography}
\end{document}